\documentclass[letter]{aa}
\usepackage{graphicx}
\usepackage{xcolor}
\usepackage{longtable,listings}
\usepackage{txfonts}

\def\dif{\mathop{}\!\mathrm{d}}

\begin{document}

\title{Mass evolution of broad line regions to explain the luminosity variability of broad H$\alpha$ in the 
TDE ASASSN-14li}


\titlerunning{$L_{H\alpha}(t)$ in TDEs}

\author{XueGuang Zhang}
\institute{Guangxi Key Laboratory for Relativistic Astrophysics, School of Physical Science and Technology,
GuangXi University, Nanning, 530004, P. R. China \ \ \ \ \email{xgzhang@gxu.edu.cn}}

\abstract{ 
In this manuscript, an oversimplified model is proposed for the first time to explain the different variability trends in 
the observed broad H$\alpha$ emission line luminosity $L_{H\alpha}(t)$ and in the TDE model determined bolometric luminosity 
$L_{bol}(t)$ in the known TDE ASASSN-14li. Considering broad emission line regions (BLRs) lying into central 
accretion disk related to accreted materials onto central black hole in TDEs, mass evolution $M_{BLRs}(t)$ of central BLRs 
can be determined by the maximum mass $M_{BLRs,0}$ of central BLRs minus the corresponding accreted mass in a TDE. Meanwhile, 
through the simple linear dependence of Broad Balmer emission line luminosity on mass of BLRs, the mass evolution $M_{BLRs}(t)$ 
of central BLRs can be applied to describe the observed $L_{H\alpha}(t)$. Although the proposed model is oversimplified 
with only one free model parameter $M_{BLRs,0}$, the model with $M_{BLRs,0}\sim0.02{\rm M_\odot}$ can be applied to well 
describe the observed $L_{H\alpha}(t)$ in the TDE ASASSN-14li. Meanwhile, the oversimplified model can also be applied to 
roughly describe the observed $L_{H\alpha}(t)$ in the TDE ASASSN-14ae. The reasonable descriptions to the observed 
$L_{H\alpha}(t)$ in ASASSN-14li and ASASSN-14ae indicate the oversimplified model with only one free model parameter is 
probably efficient enough to describe mass evolutions of $M_{BLRs}$ related to central accreted debris in TDEs.
}


\maketitle

\section{Introduction}

	ASASSN-14li ($z=0.0206$) as a well-known tidal disruption event (TDE) candidate has been firstly reported and discussed 
in \citet{ho16} through its six-month multi-band variability. And then, \citet{mg19} have shown the best descriptions to the 
long-term multi-band photometric light curves through the theoretical TDE model \citep{gr13, gm14} (the public MOSFIT code) 
applied with a 0.2${\rm M_\odot}$ main-sequence star tidally disrupted by the central supermassive black hole (SMBH) with BH 
mass about ${\rm 9\times10^6M_\odot}$, to provide stable evidence to support ASASSN-14li as a normal optical TDE.  
More recent descriptions and discussions on the MOSFIT can be found in \citet{nl22}. Meanwhile, besides the TDE model expected 
long-term multi-band photometric variability shown in \citet{ho16, mg19}, ASASSN-14li has its long-term variability of broad 
Balmer emission line luminosity $L_{H\alpha}(t)$ shown in \citet{ho16}. On study of long-term variability of $L_{H\alpha}(t)$ 
in ASASSN-14li should probably provide valuable clues on evolutions of broad emission line regions (BLRs) related to accreted 
TDE debris.

	For the TDE ASASSN-14li, based on the TDE model determined variability of bolometric luminosity (also the corresponding 
variability of accretion rate) as shown in Fig.~8\ in \citet{mg19}, the time dependent optical continuum luminosity 
$L_{5100}(t)$ at rest wavelength 5100\AA~ can be determined, after simply accepted the bolometric luminosity linearly scaled 
to the optical continuum luminosity as in normal broad line AGN as discussed in \citet{rg06, db20, nh20, sf24} for BH accreting 
systems. Meanwhile, if also simply accepted physical properties of BLRs related to TDEs debris similar as those of the normal 
BLRs in broad line AGN for recombination broad Balmer emission lines, then after considering the linear dependence of 
$L_{H\alpha}$ for the broad Balmer emission lines from central BLRs on optical continuum luminosity $L_{5100}$ at 
rest wavelength 5100\AA~ of normal quasars as shown in \citet{gh05} (the linear dependence can also be checked in the database 
of SDSS quasars in \citealt{sh11}), there should be similar variability trend in the time dependent $L_{H\alpha}(t)$ as the 
trend in the time dependent bolometric luminosity $L_{bol}(t)$ in the TDE ASASSN-14li. However, as shown in Fig.~\ref{lmc} which 
will be described in the following section, the trend of time dependent $L_{H\alpha}(t)$ is very different from the trend of time 
dependent $L_{bol}(t)$ in the TDE ASASSN-14li, indicating evolution properties of the central BLRs related to TDE debris should 
be considered, which is the main objective of this manuscript.

	Furthermore, for the BLRs related to accreted debris in TDEs, materials in the BLRs are totally from the tidally 
disrupted stars. Considering evolutions of central BLRs related to TDEs debris, especially mass evolutions of central BLRs, 
to explain $L_{H\alpha}(t)$ in the TDE ASASSN-14li, further clues should be provided on mass of the central BLRs in ASASSN-14li, 
as discussed in the following section. The manuscript is organized as follows. Section 2 presents the main hypotheses and main 
results to describe the observed $L_{H\alpha}(t)$ in the TDE ASASSN-14li. Section 3 presents the necessary discussions. Section 
4 gives our final summary and conclusions. And in this manuscript, we have adopted the cosmological parameters of 
$H_{0}=70{\rm km\cdot s}^{-1}{\rm Mpc}^{-1}$, $\Omega_{\Lambda}=0.7$ and $\Omega_{\rm m}=0.3$.

\section{Main Hypotheses and main results} 

\begin{figure}
\centering\includegraphics[width = 8cm,height=7.5cm]{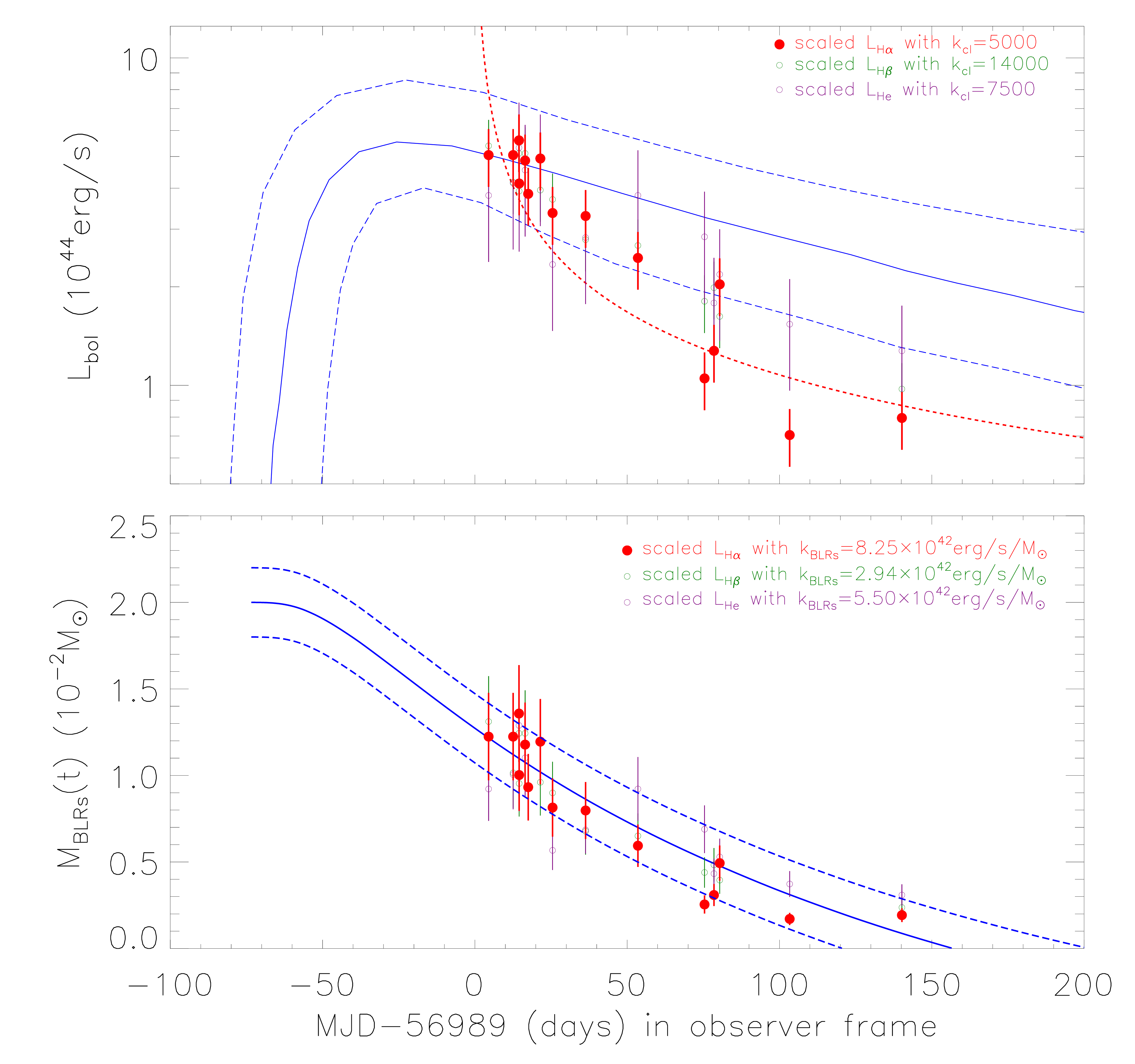}
\caption{Top panel shows the MOSFIT determined time dependent bolometric luminosity $L_{bol}(t)=\eta\dot{M_a}(t) c^2$ (solid blue 
line) and the corresponding confidence bands (dashed blue lines) in the TDE ASASSN-14li. As the shown legend in top 
right corner, solid circles plus error bars in red show the scaled $L_{H\alpha}(t)\times k_{cl}$ with $k_{cl}=5000$, open circles 
plus error bars in dark green and in purple show the corresponding results with $k_{cl}=14000$ and $k_{cl}=7500$ applied to the 
observed $L_{H\beta}(t)$ and $L_{He}(t)$ in the TDE ASASSN-14li. Bottom panel shows the determined $M_{BLRs}(t)$ (solid blue 
line) applied to describe the $L_{H\alpha}(t)=k_{BLRs}\times M_{BLRs}(t)$ in the TDE ASASSN-14li, with dashed blue lines for the 
determined confidence bands after accepted the uncertainties of the $M_{BLRs,0}$. As the shown legend in top right 
corner, solid circles plus error bars in red show the scaled $L_{H\alpha}(t)/k_{BLR}$ with 
$k_{BLRs}={\rm 8.25\times10^{42}erg/s/M_\odot}$, open circles plus error bars in dark green and in purple show the corresponding 
results with $k_{BLRs}={\rm 2.94\times10^{42}erg/s/M_\odot}$ and $k_{BLRs}={\rm 5.50\times10^{42}erg/s/M_\odot}$ applied to the
observed $L_{H\beta}(t)$ and $L_{He}(t)$ in the TDE ASASSN-14li.}
\label{lmc}
\end{figure}

	For the TDE ASASSN-14li, top panel of Fig.~\ref{lmc} shows the observed $L_{H\alpha}(t)$, similar as the results in 
Fig.~7\ in \citet{ho16}. The time dependent $L_{H\alpha}(t)$ can be described by 
\begin{equation}
	L_{H\alpha}(t)~\propto~t^{-0.638\pm0.046}
\end{equation}, 
as the shown dashed red line in the top panel of Fig.~\ref{lmc}. Meanwhile, based on the reported TDE model parameters in 
\citet{mg19}, the MOSFIT for the standard theoretical TDE model \citep{gr13, gm14} can be applied to determine the time 
dependent bolometric luminosity $L_{bol}(t)$ shown as blue lines in the top panel of Fig.~\ref{lmc}, same as the results 
in Fig.~8 in \citet{mg19} for the TDE ASASSN-14li. It is clear that the variability trend in the observed time dependent 
$L_{H\alpha}(t)$ is very different from the trend in the time dependent $L_{bol}(t)$ in the TDE ASASSN-14li. Here, in order 
to compare the $L_{bol}(t)$ with the observed $L_{H\alpha}(t)$ in the top panel of Fig.~\ref{lmc}, a scaled factor 
$k_{cl}=5000$ is applied, leading the scaled $L_{H\alpha}(t)\times k_{cl}$ to have similar magnitudes as $L_{bol}(t)$ 
around MJD-56989=10days. And, the scaled factor $k_{cl}$ has no other special physical meanings, it is only applied to 
lead the scaled $L_{H\alpha}(t)\times k_{cl}$ to be clearly shown in the top panel of Fig.~\ref{lmc}.

	Not similar as the commonly steady BLRs in normal broad line AGN leading to linear correlation between bolometric 
luminosity and broad line luminosity, there should be mass evolutions of the central BLRs related to central accreted debris 
in TDEs, which could be applied to explain the results in the top panel of Fig.~\ref{lmc} in the TDE ASASSN-14li. Then, the 
following three main hypotheses have been accepted.

	First, the upper limit of the total mass $M_{BLRs}$ of central BLRs related to TDEs debris is the total accreted mass of 
the TDEs. For the TDE ASASSN-14li, the total accreted mass is about 40\% of the mass of the tidally disrupted main sequence star 
with stellar mass of 0.2${\rm M_\odot}$ as reported in \citet{mg19}, leading the upper limit of $M_{BLRs}$ to be about 
0.08${\rm M_\odot}$. Without any other sources of broad Balmer emission line materials in TDEs, especially TDEs in quiescent 
galaxies similar as the TDE ASASSN-14li, the hypothesis can be reasonably accepted.

	Second, the broad Balmer emission line luminosity is simply scaled to the mass $M_{BLRs}$ of central BLRs related to TDEs 
debris. As discussed in the classical textbook of Astrophysics of Gaseous Nebulae and Active Galactic Nuclei \citep{of06}, through 
the $M_{BLRs}$, the broad Balmer emission line (here, broad H$\alpha$) luminosity from the BLRs can be described by 
\begin{equation}
\begin{split}
&L_{H\alpha}~=~n_e\times n_p\times\alpha_{H\alpha}^{eff}\times h\times\nu_{H\alpha}\times V_{BLRs}\times\epsilon \\
&M_{BLRs}~\sim~n_p\times M_{H}\times V_{BLRs}\times\epsilon \\
&L_{H\alpha}~=~\frac{1}{M_H}\times n_e\times\alpha_{H\alpha}^{eff}\times h\times\nu_{H\alpha}\times M_{BLRs}\\
	&\ \ \ \ \ \ \ \ \ =~k_{BLRs}\times M_{BLRs}
\end{split}
\end{equation}
with $n_e$ and $n_p$ as electron and proton density, $\alpha_{H\alpha}^{eff}$ as the effective recombination coefficient of 
H$\alpha$ emission line, $h$ as the Planck constant, $\nu_{H\alpha}$ as the emission frequency of H$\alpha$, $M_H$ 
as the proton mass, $V_{BLRs}$ as the total volume of the BLRs and $\epsilon$ as the filling factor of the materials in the BLRs. 
If simply accepted constant $n_e$ and $\alpha_{H\alpha}^{eff}$ in the BLRs related to TDEs debris, variability of the broad 
Balmer emission line luminosity should simply depend on time dependent $M_{BLRs}(t)$ (the mass evolution of central BLRs). As 
discussed in \citet{gm14} that electron density $n_e$ decreases with radius $r$ more slowly than $r^{-4}$ in physical 
environments around TDEs, therefore, constant $n_e$ can be reasonably accepted. Meanwhile, further and detailed discussions 
on few effects of temperature dependent $\alpha_{H\alpha}^{eff}$ can be found in the following section. Here, one 
point should be noted. As shown in the first sub-equation above, the parameters of $n_e$ and $\alpha_{H\alpha}^{eff}$ are 
coupled. Therefore, we do not set individual values for the parameters, but select the parameter $k_{BLRs}$ to represent the 
product of these parameters.

	Third, considering the central BLRs lying into the central accretion disk (the BLRs as part of the central accretion 
disk), the BH accreting mass should lead to decrement of the BLRs mass $M_{BLRs}$. In other words, the accreting process leads 
$M_{BLRs}(t)$ to decrease over time, which can be simply described by  
\begin{equation}
	M_{BLRs}(t)~=~M_{BLRs,0}~-~\int_{t'~\le~t}\dot{M_{a}}(t')\dif t'
\end{equation}
with $\dot{M_{a}}(t)$ as the physical accretion rates of TDEs, and $M_{BLRs,0}$ as the maximum mass of central BLRs before 
mass loss of the central BLRs. Due to the bolometric luminosity $L_{bol}(t)=\eta\times \dot{M_{a}}(t) c^2$ ($\eta$ as the 
energy transfer efficiency, $c$ as the speed of light), the shown $L_{bol}(t)$ in the top panel of Fig.~\ref{lmc} also can 
be accepted as the variability pattern of the $\dot{M_{a}}(t)$ in the TDE ASASSN-14li with the MOSFIT determined $\eta=20\%$. 
For the TDE ASASSN-14li, the $L_{bol}(t)$ (the corresponding $\dot{M_{a}}(t)$) has been determined by the MOSFIT 
and reported in \citet{mg19}. Mainly due to both the high quality $L_{H\alpha}(t)$ and the well determined TDE model expected 
$L_{bol}(t)$ (or TDE model expected $\dot{M_{a}}(t)$) reported in the literature, the ASASSN-14li is collected as the subject 
of the manuscript. Moreover, disk-like BLRs related to TDEs debris (or BLRs related to TDEs debris lying into central accretion 
disks in TDEs) have been discussed and reported in some optical TDEs candidates with double-peaked (or very asymmetric) broad 
Balmer emission lines, such as the discussions in SDSS J0159 in \citet{zh21}, ASASSN-14ae in \citet{ho14}, PTF09djl in 
\citet{lz17}, ASASSN-14li in \citet{cl18}, PS18kh in \citet{ht19}, AT2018hyz in \citet{sn20, hf20}, AT2020zso in \citet{wn22}, 
AT2019qiz in \citet{sl23}, SDSS J1605 in \citet{zh24}, etc.. Therefore, the third hypothesis on BLRs lying into central 
accretion disks in TDEs can be reasonably accepted.

	Based on the three simple hypotheses above with only one free model parameter $M_{BLRs,0}$, the observed time dependent 
broad Balmer emission line luminosity $L_{H\alpha}(t)=k_{BLRs}\times M_{BLRs}(t)$ can be determined by the time dependent 
$M_{BLRs}(t)$. Now, through Levenberg-Marquardt least-squares minimization technique \citep{mc09}, bottom panel of Fig.~\ref{lmc} 
shows the best descriptions to the $L_{H\alpha}(t)$ by linearly scaled $M_{BLRs}(t)$ with the determined 
$M_{BLRs,0}=0.02\pm0.002{\rm M_\odot}$ (about 25\% of the total accreted mass) and $k_{BLRs}\sim8.25\times10^{42}{\rm erg/s/M_\odot}$ 
in the TDE ASASSN-14li.

	Therefore, the oversimplified model with only one free model parameter $M_{BLRs,0}$ (the factor $k_{BLRs}$ only applied 
to change the $L_{H\alpha}(t)$ in units of ${\rm erg/s}$ to $M_{BLRs}(t)$ in units of ${\rm M_\odot}$) can be applied to describe 
the observed $L_{H\alpha}(t)$ in the TDE ASASSN-14li. Meanwhile, considering the linear correlation between $L_{H\alpha}(t)$ and 
$L_{H\beta}(t)$ (H$\beta$ line luminosity) ($L_{He}(t)$, He~{\sc ii} line luminosity) as shown in \citet{ho16}, the corresponding 
results for the $L_{H\beta}(t)$\footnote{There is a data point at MJD=57006 with luminosity ratio 1.36 (very smaller than the 
common value 2.8 or 3.1) of broad H$\alpha$ to broad H$\beta$. Therefore, the data point at MJD=57006 has been removed from the 
$L_{H\beta}(t)$.} ($L_{He}(t)$) in the TDE ASASSN-14li can be obtained by accepted $k_{cl}=14000$ and 
$k_{BLRs}=2.94\times10^{42}{\rm erg/s/M_\odot}$ ($k_{cl}=7500$ and $k_{BLRs}=5.5\times10^{42}{\rm erg/s/M_\odot}$), and also shown 
in Fig.~\ref{lmc}.

\begin{figure}
\centering\includegraphics[width = 8cm,height=5cm]{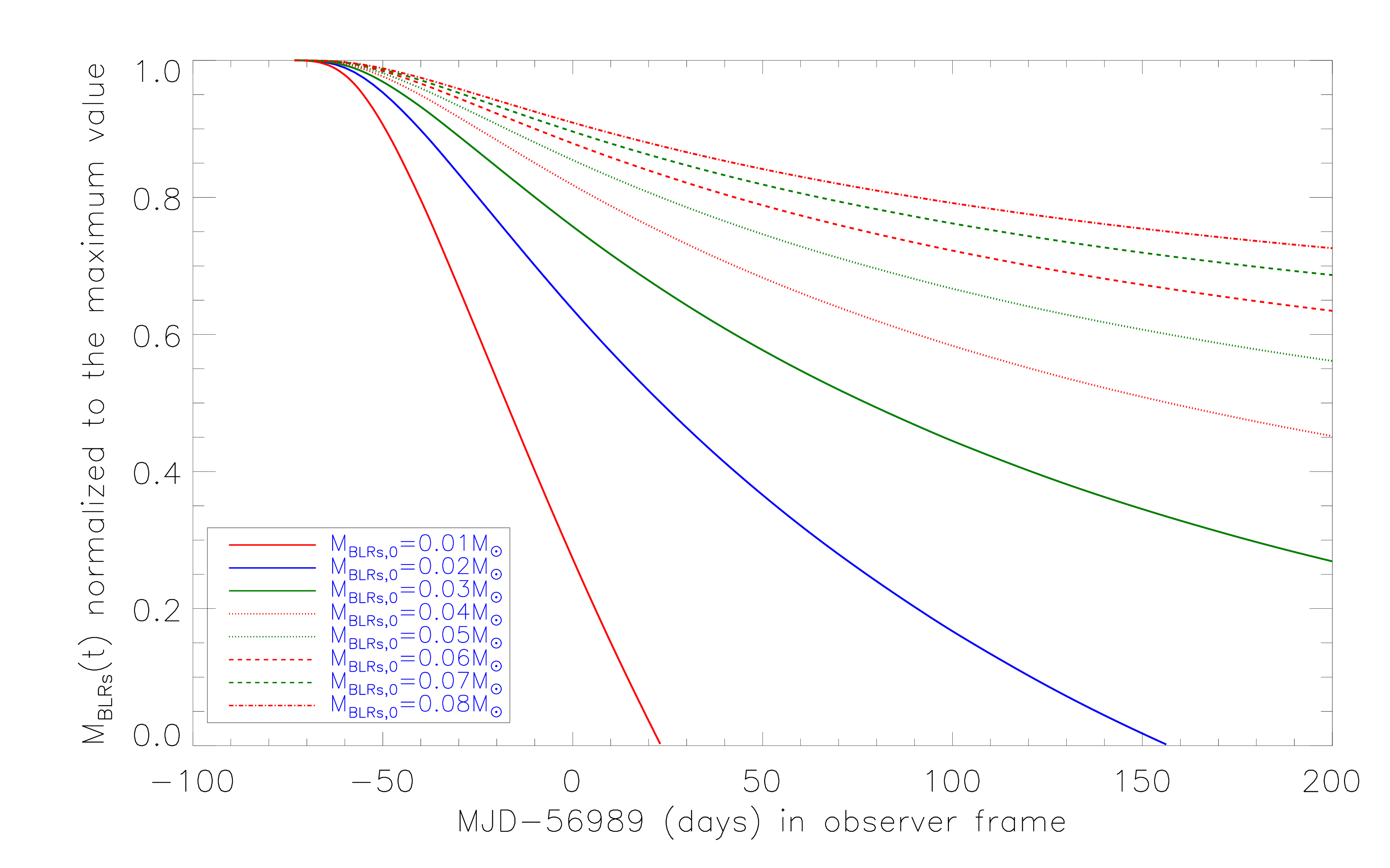}
\caption{Effects of $M_{BLRs,0}$ on $M_{BLRs}(t)$. In order to show clear comparisons, each $M_{BLRs}(t)$ has been normalized 
to its maximum value. Different line styles in different colors represent the results for different $M_{BLRs,0}$ as shown in the 
Legend in bottom-left corner. The solid blue line shows the same $M_{BLRs}(t)$ as the one shown in the bottom panel of 
Fig.~\ref{lmc}.}
\label{l2c}
\end{figure}

\section{Necessary Discussions}

	First, effects of different $M_{BLRs,0}$ are discussed. Based on the oversimplified model above applied in the TDE 
ASASSN-14li, some $M_{BLRs}(t)$ are shown in Fig.~\ref{l2c} with different values of $M_{BLRs,0}$ from 0.01 to 0.08 (the total 
accreted mass in the TDE ASASSN-14li). It is clear that different values of $M_{BLRs,0}$ can lead to different dependence 
of $M_{BLRs}(t)$ over time. Moreover, the shown results in Fig.~\ref{l2c} can also indicate that there should be different 
variability trends of broad Balmer emission line luminosity over time in different TDEs candidates, such as larger $M_{BLRs,0}$ 
relative to the total accreted masses leading to very flatter time dependent $L_{H\alpha}(t)$.

	Second, it is necessary to check whether the factor $k_{BLRs}=8.25\times10^{42}{\rm erg/s/M_\odot}$ applied in the 
TDE ASASSN-14li is reasonable. As discussed in the classical textbook of Astrophysics of Gaseous Nebulae and Active Galactic 
Nuclei \citep{of06}, the total ionized mass of BLRs in common Seyfert galaxies should be about 40${\rm M_\odot}$ for 
$L_{H\alpha, Book}\sim10^{43}{\rm erg/s}$ with $n_e\sim10^9{\rm cm^{-3}}$. Now in the TDE ASASSN-14li, the maximum 
$L_{H\alpha,0}\sim10^{41}{\rm erg/s}$ leads the estimated $M_{BLRs,0}$ to be about 
$40{\rm M_\odot}\frac{L_{H\alpha,0}}{L_{H\alpha, Book}}\sim0.04{\rm M_\odot}$, accepted $n_e\sim10^9{\rm cm^{-3}}$, very 
consistent with the determined $M_{BLRs,0}=0.02{\rm M_\odot}$. Therefore, the applied factor $k_{BLRs}$ is reasonable.

\begin{figure}
\centering\includegraphics[width = 8cm,height=7.5cm]{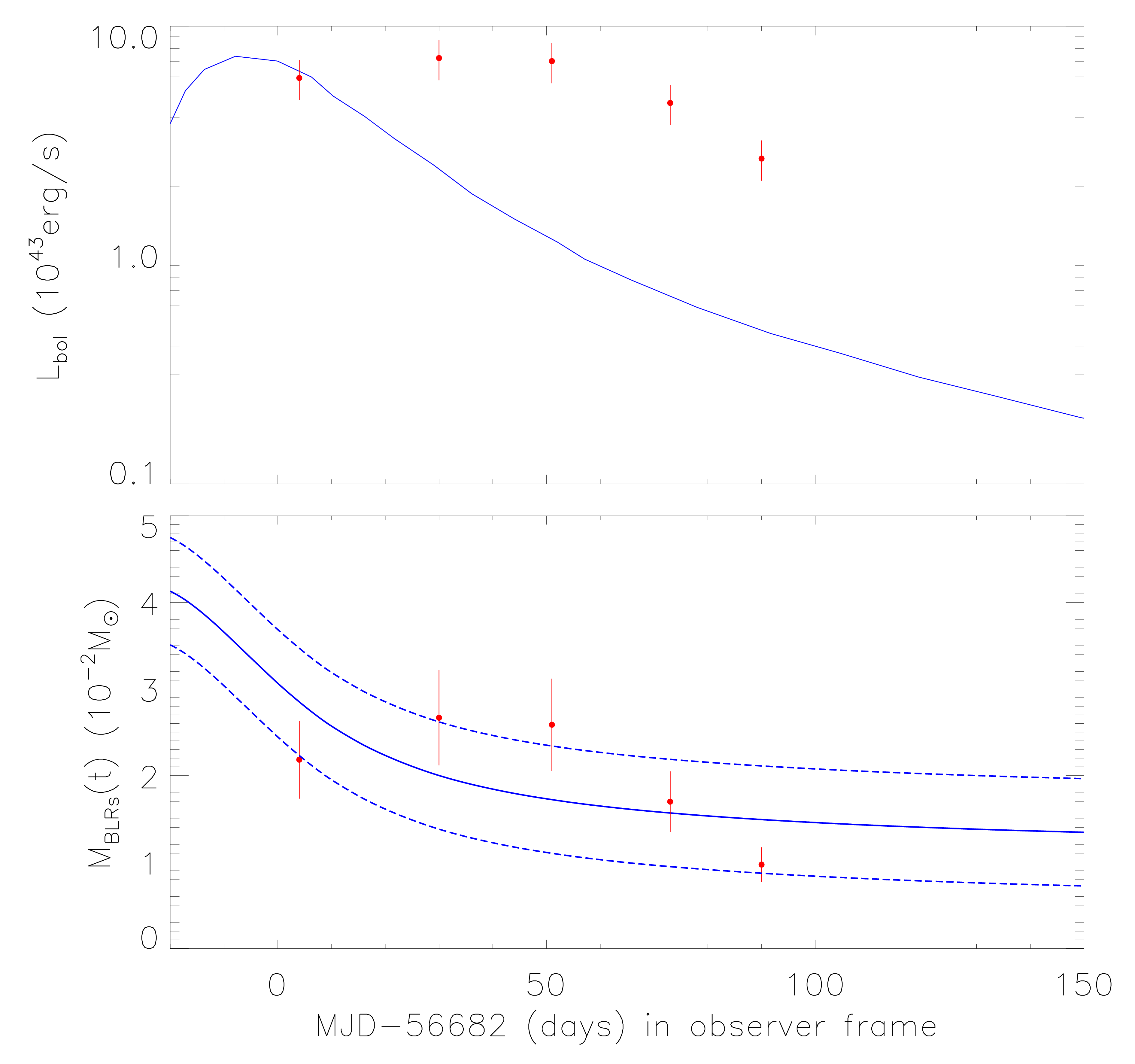}
\caption{Similar as the results in Fig.~\ref{lmc}, but for the TDE ASASSN-14ae.}
\label{l3c}
\end{figure}

	Third, it is necessary to check whether the oversimplified model can be applied to explain variability of broad Balmer 
emission lines in any other TDEs candidates. Among the reported more than 150 TDEs candidates, such as the candidates reported 
in \citet{gm06, ce12, wz17, wy18, gs21, sg21, vg21, zs22, zh22, yr23, zh24b}, etc., ASASSN-14ae is one another TDE with reported 
broad H$\alpha$ luminosity in \citet{ho14}, although only five data points in $L_{H\alpha}(t)$. Then, the MOSFIT determined 
$L_{bol}(t)$ as shown in \citet{mg19} and the observed $L_{H\alpha}(t)$ from \citet{ho14} are shown in the top panel of 
Fig.~\ref{l3c} in the TDE ASASSN-14ae. The variability trend of the observed $L_{H\alpha}(t)$ is very different from the trend 
of the MOSFIT determined $L_{bol}(t)$ in the TDE ASASSN-14ae. Then, the oversimplified model discussed above can be applied to 
describe the $L_{H\alpha}(t)$ in the TDE ASASSN-14ae, as shown in the bottom panel of Fig.~\ref{l3c} with 
$M_{BLRs,0}=0.04\pm0.004{\rm M_\odot}$ (the MOSFIT determined total accreted mass about $0.04\pm0.01{\rm M_\odot}$). Similar as 
the expected results in Fig.~\ref{l2c}, the flatter $L_{H\alpha}(t)$ than $L_{bol}(t)$ leads to the determined $M_{BLRs,0}$ near 
to the total accreted mass in the TDE ASASSN-14ae. Here, we should note that there are only five data points in 
$L_{H\alpha}(t)$ in the TDE ASASSN-14ae, it is hard to give clear statements that the oversimplified model can be applied to 
well describe the observed $L_{H\alpha}(t)$. However, considering almost all the five data points are lying within the 
confidence bands as shown in bottom panel of Fig.~\ref{l3c}, we can give clear statements that the descriptions determined by 
our oversimplified model can be roughly accepted in the TDE ASASSN-14ae.

	Fourth, in the oversimplified model, there are no considerations of effects of the temperature ($T$) dependent effective 
recombination coefficient $\alpha_{H\alpha}^{eff}(T)$. Considering the MOSFIT determined time dependent photosphere temperature 
$T(t)$, as shown in Fig.~8 in \citet{mg19}, the environment temperature can be changed from $6.76\times10^4$K at MJD-56989=10days 
to $1.08\times10^5$K at MJD-56989=145days in the TDE ASASSN-14li, leading to only 25\% changes in $\alpha_{H\alpha}^{eff}(T)$. 
In other words, the variability of the environment temperatures can lead to only 25\% changes in the observed $L_{H\alpha}(t)$ 
in the TDE ASASSN-14li. Comparing with uncertainty of $L_{H\alpha}$ about 20\% and with the ratio 8 of the $L_{H\alpha}(t)$ at 
MJD-56989=10days to the $L_{H\alpha}(t)$ at MJD-56989=145days in the TDE ASASSN-14li, the effects of $\alpha_{H\alpha}^{eff}(T)$ 
can be totally ignored in ASASSN-14li. Meanwhile, through the MOSFIT determined results in the TDE ASASSN-14ae, the environment 
temperature can be changed from $1.80\times10^4$K at MJD-56682=4days to $2.15\times10^4$K at MJD-56682=90days, leading to only 
9\% changes in $\alpha_{H\alpha}^{eff}(T)$. Comparing with the ratio about 2 of the $L_{H\alpha}(t)$ at MJD-56682=4days to the 
$L_{H\alpha}(t)$ at MJD-56682=90days in the TDE ASASSN-14ae, the effects of $\alpha_{H\alpha}^{eff}(T)$ can also be ignored in 
ASASSN-14ae.

	Before proceeding further, we have noted that there are many TDEs candidates of which $L_{H\alpha}(t)$ have been reported 
in the literature, such as the discussed and reported spectroscopic properties of a small sample of TDEs in \citet{cl22}. However, 
we mainly consider the TDEs of which TDE model expected $\dot{M_a}(t)$ (or the corresponding $L_{bol}(t)$) has been determined 
and reported in the literature, which will be applied in the Equation 3 in this manuscript. To determined $\dot{M_a}(t)$ through 
applications of TDE model to describe long-term photometric variability is beyond the scope of this manuscript. Therefore, only 
ASASSN-14li and ASASSN-14ae are mainly considered and discussed, because their $L_{bol}(t)$ ($=\eta\dot{M_a}(t)c^2$) have been 
reported in \citet{mg19}. For the reported TDEs candidates in \citet{cl22} and the references therein, there are apparent 
$L_{H\alpha}(t)$ but no clear information of TDE model determined $L_{bol}(t)$ (or $\dot{M_a}(t))$. Therefore, the TDEs candidates 
listed and discussed in \citet{cl22} are not considered in this manuscript.

	Although the discussed model above is oversimplified, the model can be applied to describe the observed time dependent 
luminosity variability of broad Balmer emission lines from central BLRs related to TDEs debris in the known TDEs ASASSN-14li 
and ASASSN-14ae, indicating the oversimplified model with only one free model parameter $M_{BLRs,0}$ is efficient enough to 
some extent. In the near future, to test the oversimplified model in more TDEs should provide further clues to support or to 
be against the oversimplified model. Certainly, in our oversimplified model, there are no time delays between formations of 
BLRs and formations of central accretion disks related to TDEs debris. In the near future, if clear clues can be found in any 
TDE candidate for such times delays ($t_d$) between TDE model determined $L_{bol}(t)$ and high quality observed $L_{H\alpha}(t)$, 
the $M_{BLRs}(t)$ could be probably improved to be
\begin{equation}
	M_{BLRs}(t)~=~M_{BLRs,0}~-~\int_{t'~\le~t~+~t_d}\dot{M_{a}}(t')\dif t'
\end{equation},
which will lead to more flexible results on expected $L_{bol}(t)$ in TDEs. Unfortunately, we currently do not have any clues for 
such delays.

\section{Summary and Conclusions}

	The final summary and main conclusions are as follows. 
\begin{itemize}	
\item Based on the TDE model determined time dependent bolometric luminosity $L_{bol}(t)$ and the observed time dependent broad 
H$\alpha$ emission line luminosity $L_{H\alpha}(t)$ in the known TDE ASASSN-14li, there are not consistent variability trends in the 
$L_{bol}(t)$ and in the $L_{H\alpha}(t)$, indicating different properties of the BLRs related to TDEs debris from those of the 
steady BLRs in normal broad line AGN.
\item In order to explain the observed $L_{H\alpha}(t)$ in the TDE ASASSN-14li, an oversimplified model is proposed after 
considering the BLRs related to TDEs debris accreted by central BH, leading to decrement of the mass of BLRs $M_{BLRs}(t)$ over 
time. Then, the observed $L_{H\alpha}(t)$ can be linearly scaled by the $k_{BLRs}\times M_{BLRs}(t)$.
\item Based on the oversimplified model with only one free model parameter $M_{BLRs,0}$ (the expected maximum mass of BLRs before 
mass loss of the BLRs), the observed $L_{H\alpha}(t)$ can be well described in the TDE ASASSN-14li with the determined parameter 
$M_{BLRs,0}~\sim~0.02\pm0.002{\rm M_\odot}$. 
\item Meanwhile, the oversimplified model can also be applied to roughly describe the observed $L_{H\alpha}(t)$ in the TDE 
	ASASSN-14ae.
\item Furthermore, after considering different values of $M_{BLRs,0}$ and probably time delays between formations of BLRs and 
formations of accretion disks in the oversimplified model, more flexible results on $L_{H\alpha}(t)$ could be expected in TDEs.
\end{itemize}

\begin{acknowledgements}
Zhang gratefully acknowledge the anonymous referee for giving us constructive comments and suggestions to greatly 
improve the paper. Zhang gratefully thanks the kind financial support from GuangXi University and the kind grant support from 
NSFC-12173020 and NSFC-12373014. This manuscript has made use of the public TDEFIT (\url{https://github.com/guillochon/tdefit}), 
MOSFIT(\url{https://github.com/guillochon/mosfit}), and the MPFIT package (\url{http://cow.physics.wisc.edu/~craigm/idl/idl.html}).
\end{acknowledgements}


\label{lastpage}
\end{document}